\documentstyle[11pt]{article}

\def\l*{$L_*$\/}

\def\kms{\rm ~km~s^{-1}}

\def\l*{$L_{*}$}

\def\gsim{ \lower .75ex \hbox{$\sim$} \llap{\raise .27ex \hbox{$>$}} }
\def\lsim{ \lower .75ex \hbox{$\sim$} \llap{\raise .27ex \hbox{$<$}} }

\def\spose#1{\hbox to 0pt{#1\hss}}
\def\simlt{\mathrel{\spose{\lower 3pt\hbox{$\mathchar"218$}}
     \raise 2.0pt\hbox{$\mathchar"13C$}}}
\def\simgt{\mathrel{\spose{\lower 3pt\hbox{$\mathchar"218$}}
'     \raise 2.0pt\hbox{$\mathchar"13E$}}}

\font\titlefont=cmss17

\input{epsf}

\begin{document}

\noindent{\titlefont Gone with the wind:}
\noindent{\titlefont The origin of S0 galaxies in clusters}

\bigskip

\noindent{\bf Vicent Quilis, Ben Moore \& Richard Bower}
\footnote{vicent.quilis@durham.a.uk, ben.moore@durham.ac.uk, 
r.g.bower@durham.ac.uk}

\noindent{Department of Physics, Durham University, 
South Road, Durham, DH1 3LE, UK}

\bigskip
\bigskip


\noindent{\bf  
We present the first 3-dimensional high resolution hydro-dynamical
simulations of the interaction between the hot ionised intra-cluster
medium and the cold interstellar medium of spiral galaxies.  Ram pressure
and turbulent/viscous stripping removes 100\% of the atomic hydrogen
content of luminous galaxies like the Milky Way within 100 million
years. These mechanisms naturally account for the morphology of S0
galaxies, the rapid truncation of star formation implied by spectroscopic
observations, as well as a host of observational data on the HI morphology
of galaxies in clusters.}

\bigskip

Crucial observational evidence for the hierarchical formation of
structure in the universe is the dramatic evolution of galactic morphologies 
in dense environments over the past 5 billion years (\ref{BO},\ref{D97}). 
The key puzzle that remains to be solved is the origin of the large 
population of lenticular (S0) galaxies found in nearby clusters 
(\ref{sandage70},\ref{D80}).  These featureless disky galaxies contain 
no atomic gas and show no signs of recent star-formation (\ref{sandage70}).

The Hubble Space Telescope revolutionised our view of the universe by 
revealing that distant galaxies appeared different from the local 
population.  In contrast to local clusters, high resolution imaging of 
distant clusters led to the spectacular finding that young clusters of 
galaxies are filled with spiral galaxies (\ref{D97},\ref{D80},\ref{C98})
and contain almost no lenticular (S0) galaxies, whereas the ratio of luminous
ellipticals to lenticulars (S0) increases by a factor of five between a 
redshift z=0.5 and the present-day (\ref{D97}). S0's can be characterised 
by their thick featureless disks that show no evidence
for recent star-formation and the increase in their population appears to be
countered by a similar decrease in the number of luminous late-type spirals in
clusters.  The data suggest that a transformation between these galaxy 
types is taking place as a direct consequence of the cluster environment.

Three mechanisms have been proposed that can lead to morphological
transformation between galaxy classes.  Mergers will transform disks to
spheroidals 
(\ref{toomre},\ref{barnes}), but are only 
effective in low density
environments 
(\ref{ghigna}). Gravitational tidal interactions between cluster
galaxies can naturally account for the observed evolution of the faint end of
the luminosity function and the transformation of small disks to faint
spheroidals 
(\ref{moore}).  However, more massive bulge dominated systems are
stable to tidal disruption 
(\ref{moore99}). Although the resulting disk
thickening from tidal heating suppresses spiral features and causes a 
morphological similarity to
S0's, neither of these two processes suppress star-formation.

In addition to disk thickening, a 
mechanism that actively extinguishes star-formation is crucial since the
stellar populations of S0 galaxies are old and their spectra indicate that
star-formation was {\it abruptly} halted several billion years ago
(\ref{DG},\ref{CS},\ref{poggianti}).  A slow
decline in the star formation rate, such as expected from the exhaustion of a
reservoir of cold gas, is unable to explain the strongly enhanced Hydrogen
absorption lines seen in many distant cluster galaxies.

A candidate mechanism was proposed by Gunn \& Gott 
(\ref{GG}) over two decades
ago.  Their simple force-balance 
estimates suggested that the motion of galaxies
through the hot ionised intra-cluster medium (ICM) creates a 
``ram-pressure'' that
could potentially strip away significant amounts of gas from disks. However, a
full description of this mechanism must include complex turbulent and viscous
stripping 
(\ref{nulsen}) at the interface of the cold and hot gaseous components
as well as the formation of bow-shocks in the ICM ahead of the galaxy.  
Although these processes have 
been cited over 1000 times in the literature, their
effectiveness and efficiencies have received little theoretical investigation.

The ram pressure is proportional to $\rho_{_{ICM}} v_{gal}^2$, therefore the
infalling galaxy suffers most gas loss at its pericentric passage, where its
velocity can be as large as $3000 \kms$ and the intra-cluster gas density
approaches $3000 \ h_{50}^{1/2} {\rm atoms} \ m^{-3}$. 
Gunn \& Gott's 
(\ref{GG}) order of magnitude estimates suggest
that the typical gas disk would be stripped down to $\sim 5$ kpc, a radius 
confirmed by
smoothed particle hydro-dynamic simulations that follow just the ram pressure
process 
(\ref{abadi}). However, this leaves 50\% of the original HI confined in
the disk which would continue to 
form stars for several billion years. Although
this is a significant reduction 
in the star formation rate, it does not explain
the absence of any recent star formation in cluster S0 galaxies.

We have performed the first high resolution three dimensional numerical
simulations of these hydro-dynamical processes to accurately address the
efficiency of stripping and the timescale on which it occurs.  Previous 
work has
been either in 2D with spherical galaxy models 
(\ref{gaetz},\ref{balsara}), or in 3D
using a code that could not model viscosity and turbulence 
(\ref{abadi}). 
Our parallel computer code (\ref{quilis})
uses {\it high-resolution shock-capturing} techniques
to follow the fluid dynamic equations allowing us to observe
the full complexity of the ram-pressure and turbulent/viscous stripping
processes. We can also
follow the shocks that penetrate the interstellar medium
(ISM) and ICM and the thermo-dynamical evolution of the
ICM and stripped galactic gas. Because it is fully three dimensional, our code
is not forced to preserve the cylindrical symmetry of the galaxy 
(\ref{balsara}).

We construct a self-consistent equilibrium model galaxy with stellar 
disk and bulge
components designed to resemble a luminous spiral similar to
the Milky Way or Andromeda. The stars are embedded within a
dark matter halo constructed  such that the total
rotational velocity of the disk is a constant $220 \kms$ (\ref{galmod}).
The real ISM is complex gaseous medium formed by a cold diffuse HI component   
and dense molecular clouds (MC) with temperatures $T_{MC}\sim 10^2 \, K$. 
The diffuse gaseous disk is constructed  
by specifying the density, and velocity at each grid cell according with 
(\ref{galmod}),
and constant temperature $T_{ISM}=10^4 \, K$ -- the lower threshold
in our simulations.
The MC's 
are typically three order of magnitude denser than the HI component 
and have sizes of the order of several parsecs.  
Even though our numerical code is highly optimised and is running on state
of the art parallel hardware, the maximum resolution that we can achieve is
$\sim 100$ parsecs, therefore we cannot resolve individual MC's in our
simulations. However, MC's are so small and dense that they will remain
unaffected by the stripping processes (\ref{MC}). 
Initially, the ICM is considered as an uniform medium with constant 
density and temperature $T_{ICM}=10^8\, K$.
Both the ICM and ISM are treated as ideal fluids with adiabatic exponent 
$\gamma=5/3$. Their evolution is described by the hydrodynamic equations
that are integrated using the numerical techniques described above.
The stellar and dark matter components are evolved using an 
N-body Particle-Mesh code. All components are coupled gravitationally through 
Poisson's equations which is solved using a fast 3D FFT method.
No cooling has been considered since the 
timescales are so short (\ref{cooling}).
At our best (standard) resolution we use $512^3(256^3)$ cells across a
cubic region centered on the galaxy with length 64 kpc, thus our nominal
resolution is $\sim 100(200)$ parsecs.

We have simulated different infall geometries, velocities and ICM gas
densities and ISM structure (\ref{casos}). 
Two of these simulations are pictured in Figure 1, which show a
galaxy moving face on and nearly edge on through the core of a rich cluster
 like Coma -- $\rho_{_{ICM}}=2.6\times10^3 \, h_{50}^{1/2} 
 {\rm atoms} \, m^{-3} $ -- 
at a velocity of $2000 \kms$. 
We plot only the ICM and ISM components of the simulation to highlight the gas
dynamical processes - the stellar disk remains unperturbed. 
The outer gas disk of the infalling galaxy
is rapidly stripped away and forms trailing streams of warm HI in pressure
equilibrium with the ICM. Turbulence and viscous stripping ablate the gas disk
even further (\ref{viscous}).   

We find a rich 
structure of shocks. The most obvious is the prominent curved bow shock 
that propagates through the ICM ahead of the galaxy, heating the ICM from 
a temperature of $10^8$ to $5\times10^8$ 
degrees (\ref{merrifield},\ref{stevens}). 
A complex series of cross shocks occur in the rarefied hot medium
behind the galaxy, which may be visible as a wake of enhanced X-ray emission
(\ref{merrifield}). Although not apparent on this scale, a second
shock   is driven through the   ISM of the galaxy, raising the internal 
pressure
by over two orders of magnitude. The efficiency 
and time scale of star formation are
strong functions of the ambient ISM pressure (\ref{elmergreen}). As the 
shock propagates through the galaxy, it may promote the collapse of molecular 
clouds, briefly enhancing the galaxy's star formation rate.

Whereas previous work treated the ISM as a smooth disk of HI, in reality, the
ISM has complex multi-phase structure, filled with bubbles, shells and holes
ranging in size from a few parsecs to a kpc 
(\ref{elmergreen}). Furthermore, the
inner couple of kpc of most bright spiral galaxies are extremely deficient in
HI. The nearest and best studied galaxy is Andromeda 
which has over 100 HI holes
and a central region of radius 2 kpc devoid of 
neutral gas (\ref{andromeda}). 
Our
simulations show that this structure makes the disk much more susceptible to
viscous stripping. As the ICM streams through these holes,
it  ablates their edges
and prevents stripped gas falling back on to the
centre of the galaxy.  This is an important difference between previous work
that claimed significant replenishment from stripped material. When we 
model our
disk on Andromeda's we find that these processes lead to the removal of the
entire diffuse HI component on a timescale of 100 million years.
Even if the HI holes contain a large quantity of 
molecular hydrogen locked within MC's, 
due to their small covering factor they would not affect 
the removal of the HI component of the gaseous disk (\ref{MC}). 
Combined with galaxy harassment, this process has all
the properties required to explain the rapid transformation between spiral and
S0 galaxies seen in distant clusters. 

It is important to stress that the effectiveness of ram-pressure
stripping does not depend on the galaxy moving face-on through the
ICM.  We find that galaxies inclined twenty degrees to the direction
of motion suffer as much stripping as face on encounters. Since the
orientation of the disk remains fixed as the galaxy orbits through the
cluster, a galaxy moving edge on at first pericentric passage is 
likely 
to be inclined to the direction of motion at some other 
point of its orbit (\ref{orientation}). Only a small fraction of galaxies
will orbit with their disks aligned exactly with the orbital plane.

The timescale for stripping is very short compared to the
orbital timescale, therefore galaxies will only rarely be
observed at the moment
of stripping. Evidence for ongoing ICM/ISM interaction may
be found by observing  compression of the leading contours of disk HI 
emission, and wings or tails of HI behind the galaxy
(\ref{vollmer},\ref{ryder}).  In Figure 3 
(see also \ref{radiojet}) we compare two recent and striking 
examples of these processes with snapshots of our simulations, showing
that the observed morphologies are well reproduced.
After passage  through the cluster, disks will be either HI
deficient compared to field spirals, or they may lack neutral hydrogen
altogether. Galaxies in this latter class would subsequently be identified as
S0's. HI maps of galaxies in the Virgo and  Coma clusters 
show few galaxies with any HI,
although some galaxies in the outer parts of the cluster are extremely HI
deficient (\ref{cayatte},\ref{warmels},\ref{bravo}).  The 
efficiency of stripping 
explains why such galaxies are so rare.

Our simulations demonstrate the importance and effectiveness of the
ram-pressure and transport processes. The interaction between the ICM and ISM 
removes
the entire HI component as well as any diffuse reservoir of hot gas
within its dark matter halo. But what happens to the molecular clouds
in the infalling galaxy? These clouds are so dense and small that they 
cannot be removed
by the ram-pressure of the ICM and are not resolved in our simulation. 
To understand the fate of the molecular 
gas we need to consider the cycle of star formation in a little more detail.
In a quiescent field galaxy, molecular clouds are continually being
disrupted by the star clusters formed within them and they are subsequently 
reformed from the 
ambient diffuse HI. Within clusters, galaxies are swept clear
of HI and this cycle is broken: the disruption of the clouds is not
balanced by the condensation of new self-shielding molecular complexes.
Current models
suggest the lifetime of large molecular clouds is less than a few tens
of millions of years (\ref{elmergreen},\ref{blitz}). 
We would therefore expect the 
ram pressure to lead to a decline in star formation on the same
timescale as the disruption of the molecular clouds. This scenario predicts
that cluster S0 galaxies will not contain molecular gas (\ref{rgbnote}).

Furthermore, the pressure increase in the ISM may create a burst of 
star-formation, as the counter shock compresses
existing molecular clouds within the galaxy (\ref{fujita}).  
This would explain the prevalence of strongly enhanced Hydrogen 
absorption lines in many
distant cluster S0 galaxies, and is supported by the spectacular
H$_\alpha$ emission seen from several nearby galaxies as they are
stripped (\ref{gavazzi},\ref{kenney}).  

This mechanism leads to a
population of post-starburst S0 galaxies, preferentially populating
the central regions of rich clusters. However, ram-pressure is not the
only effect at work in clusters of galaxies. Longer timescale
processes, such as stellar evolution, tidal forces and `harassment'
tend to thicken the stellar disk and
enhance the relative importance of the galaxy's bulge. Taken together,
these processes can explain the remarkable transformation of cluster
galaxies and the dramatic evolution in the galaxy populations of dense
environments.
S0's in the field most likely form via a minor merger or the accretion
of a satellite (\ref{barnes}). This formation mechanism is markedly
different from what we are proposing in clusters, and suggests that 
there will be 
observable photometric, spectral and kinematical differences between
field and cluster S0's.

\noindent
{\bf References and notes}

\newcommand{\araa}{{\em Ann. Rev. Astron. Astrophys. }}
\newcommand{\mn}{{\em Mon. Not. R. Astr. Soc. }}
\newcommand{\apj}{{\em Astrophys. J. }}
\newcommand{\apjl}{{\em Astrophys. J. Letters }}
\newcommand{\apjs}{{\em Astrophys. J. Suppl. }}
\newcommand{\aj}{{\em Astron. J. }}
\renewcommand{\aa}{{\em Astron. Astrophys. }}
\newcommand{\aass}{{\em Astron. Astrophys. Suppl. }}
\newcommand{\ass}{{\em Astrophys. Space Sci. }}
\newcommand{\nat}{{\em Nature }}
\newcommand{\pasp}{{\em Pub. Astr. Soc. Aust. }}
\newcommand{\physrevD}{{\em Phys. Rev. D. }}

\begin{enumerate}
\item \label{BO}H. Butcher and A. Oemler, \apj {\bf 285}, 426 (1984).

\item \label{D97} A. Dressler et al., \apj {\bf 490}, 577 (1997).

\item \label{sandage70} A. Sandage, K.C. Freeman, N.R. Stokes, 
 \apj {\bf 160}, 831 (1970).

\item \label{D80} A. Dressler, \apjs  {\bf 42}, 565 (1980).

\item \label{C98} W.J. Couch, A.J. Barger, I. Smail, R.S. Ellis, 
R.M. Sharples, \apj {\bf 497}, 188 (1998).

\item \label{toomre}A. Toomre and J. Toomre,  
 \apj {\bf 178}, 623 (1972).

\item \label{barnes} J.E. Barnes and L.  Hernquist,
\apj {\bf 471}, 115 (1996).

\item \label{ghigna} S. Ghigna et al., \mn {\bf 300}, 146 (1998).

\item \label{moore}B. Moore, N. Katz, G. Lake, A. Dressler, A. Oemler,  
\nat {\bf 379}, 613 (1996).

\item \label{moore99} B. Moore, G. Lake, T. Quinn, J. Stadel, 
 \mn {\bf 304}, 465 (1999).

\item \label{DG} A. Dressler and  J.E. Gunn,  
 \apj {\bf 270}, 7 (1983).

\item \label{CS} W.J. Couch and  R.M. Sharples,
 \mn {\bf 229}, 423 (1987).

\item \label{poggianti} B.M. Poggianti et al.,  
 \apj {\bf 518}, 576 (1999).

\item \label{GG} J.E. Gunn and J.R. Gott,
 \apj {\bf 176}, 1  (1972).

\item \label{nulsen} P.E.J. Nulsen, \mn {\bf 198}, 1007 (1982).

\item \label{abadi} M.G. Abadi, B. Moore, R.G. Bower, 
\mn {\bf 308}, 947 (1999).

\item \label{gaetz} T.J. Gaetz, E.E. Salpeter, G. Shaviv,
\apj {\bf 316}, 530 (1987).

\item \label{balsara} D. Balsara, M. Livio, C.P. O'dea,  
\apj {\bf 437}, 83 (1994).

\item \label{quilis} V. Quilis, J.M$^{\mbox{a}}$. Ib\'a\~nez, D. S\'aez, 
\apj {\bf 469}, 11 (1996). 
The numerical code used in this report is a 3D Eulerian code 
on a fix cartesian grid. This code is based on
modern {\it high-resolution shock-capturing} (HRSC) techniques,  
a general denomination for a recently developed 
family of methods to solve hyperbolic systems of equations such as 
the hydro-dynamic equations.
Our code is similar to PPM (Piecewise Parabolic Method) but 
with some particular features. 
It  has  four key ingredients: i) conservative formulation,  
numerical quantities are conserved up to the numerical order of the
method, 
 ii) the reconstruction procedure, which allow 
to recover the distribution of the quantities inside the computational 
cells, iii) the Riemann solver, which solves the evolution of 
discontinuities between cell interfaces, and 
iv) the advancing in time, designed  to be consistent with the 
conservation properties. HRSC schemes have the following advantages;
they do not suffer from numerical artifacts 
such as {\it artificial viscosity},
they can resolve strong shocks extremely well -- typically in one or two
cells, strong gradients are perfectly modelled, they work very well in 
low density regions and are high-order in smooth regions of the 
flow. 

\item \label{galmod} 
We construct the galaxy following the Hernquist's 
model [L. Hernquist, \apjs {\bf 86}, 389 (1993)]. Four 
components are considered:
(i) Stellar bulge,
\begin{displaymath}
\rho_b(r)=\frac{M_b}{2\pi r_b^2}\frac{1}{r(1+\frac{r}{r_b})^3},
\end{displaymath}
with $r_b=0.5\ kpc$, $M_b=1.7\times10^{10}\ M_{\odot}$ 
. (ii) Dark matter halo,
\begin{displaymath}
\rho_b(r)=\frac{M_h}{2\pi^{3/2}}\frac{\alpha}{r_tr_h^2}
\frac{{\rm exp}(-r^2/r_t^2)}{(1+r^2/r_h^2)},
\end{displaymath}
with $r_h=3.5\ kpc$, $r_t=24.5\ kpc$, $M_h=26.5 \times10^{10}\ 
M_{\odot}$,  $\alpha=1/[1-\pi^{1/2}q\, e^{q^2}[1-erf(q)]]$
being $erf(q)$ the error function with $q=r_h/r_t$. 
 (iii) Stellar disk,
\begin{displaymath}
\rho_s(R,z)=\frac{M_s}{4\pi R_s^2 z_s}{\rm exp}(-R/R_s){\rm sech}^2(z/z_s),
\end{displaymath}
where $R_s= 3.5\ kpc$, $z_s=0.35\ kpc$, $M_s=5.6\times10^{10}\ 
M_{\odot}$.  
 (iv) Gas disk,
\begin{displaymath}
\rho_g(R,z)=\frac{M_g}{4\pi R_g^2 z_g}{\rm exp}(-R/R_g)sech^2(z/z_g),
\end{displaymath}
being $R_g= 3.5\ kpc$, $z_g=0.35\ kpc$,  $M_g=1.4
\times10^{10}\ M_{\odot}$.  The total masses of the different 
components are 
$5\times10^{10}\,M_{\odot}$, $1.7\times10^{10}\,M_{\odot}$ 
and $5\times10^9M_\odot$ for the disk of stars, bulge, and
gaseous disk, respectively. $10^5$ and $1.4\times10^5$ 
particles are used to describe the DM halo and the stellar components
respectively.

\item \label{MC} Molecular clouds are structures much smaller than the maximum
numerical resolution that we can achieve (100 parsecs). 
Therefore, they can not be 
modelled in our simulations as components of the ISM which it is described as 
exponential disk of cold HI. Nevertheless, we can conclude that MC's are not
relevant to the ram-pressure stripping suffered by the HI component due to 
their small size and high density. Several previous studies, such as A.C. Raga, 
J. Cant\'o, S. Curiel, S. Taylor, \mn, {\bf 295}, 738 (1998) (and references 
therein) justify this statement. In this paper, the authors carried out an 
analytical and numerical study of the interaction of MC's 
with winds. If we apply their conclusions to the typical 
parameters adopted here,
the clouds would remain unaltered, that is, MC's do not suffer ram-pressure
stripping by the interaction with the ICM. A second possible effect of
MC's embedded in the flow is to shield the HI component from the ICM.
This effect is also negligible due to the small covering 
factor of MC's. Their cross-sections is less than 1\% of the area of one of 
our numerical cells, thus they would act like single points in a fluid.

\item \label{cooling} The ISM is heated very efficiently by shocks
to a temperature $T_{ISM}=10^6$ in a very short timescale.
However, the cooling time for the ISM material with typical metal 
abundance is much shorter than the dynamical timescale. A good approximation
is that all the energy imparted into the ISM via shocks is immediately 
re-radiated away, possibly as $H_\alpha$ photons.
 
\item \label{casos} We have carried out a set of simulations setting
different values for the ICM parameters, orientation of the galaxy against
the wind, and the composition of the ISM. Two ICM densities have 
been considered $0.1\rho_{_{coma}}$ and 
$\rho_{_{coma}}\, 
(\rho_{_{coma}}=2.6\times10^3\,h_{50}^{1/2} atoms\,m^{-3})$.
The ICM velocities used were $1000 \, km\, s^{-1}$ and $2000 \, km\, s^{-1}$.
The orientation of the galaxies moving through the wind were varied from 
face-on to edge-on, passing through
45 and 20 degrees. We use three ISM compositions: 1) uniform 
smooth exponential
disk (\ref{galmod}). 2) The previous disk but with a central region devoid 
of diffuse HI gas with a 2 kpc radius. 3) The original exponential disk
in which ten small holes each of radius 300 parsecs are 
randomly located within a 5 kpc radius from the center, and in the same 
region the local density of the cells is randomly 
increased by a factor of two with a 50\% probability. This last case pretends
to resemble an inhomogeneous ISM.      
All of the simulations show a rapid loss of gas but the models with 
$\rho_{_{ICM}}=0.1\rho_{_{coma}}$, $v_{_{ICM}}=1000\, km\, s^{-1}$ 
and no holes, are not able to remove the bulk of this material.
Simulations with high ICM velocity and density 
but with a smooth ICM with no holes retain small HI disk with 
sizes of 3 kpc after 100 Myrs. The more realistic cases
including a non-uniform ISM exhibit massive gas losses with 
almost no HI component remaining after 100 Myrs. Only the strict edge-on cases 
are weakly affected by the stripping
processes, but this configuration for several 
orbits is expected to be quite rare.
Results of one simulation including ten small holes and inhomogeneous 
density are shown as mpeg movies in http://www.scienceonline.org.

\item \label{viscous} Ram pressure stripping removes the outer disk gas
in a timescale of 20 Myrs.
Turbulence and viscous stripping operate over the entire surface of the disk
and are effective at removing the diffuse HI even from regions of the disk
that are above the threshold for ram pressure effects. These latter processes
operate over a longer timescale and are effective at depleting the diffuse
HI from the central disk in a timescale of order of 
the crossing time for the ICM through the ISM
(see Figure 2).

\item \label{merrifield} M.R. Merrifield,
\mn {\bf 294}, 347 (1998).

\item \label{stevens} I.R. Stevens, D.M. Acreman, T.J. Ponman, 
\mn {\bf 310}, 663 (1999).

\item \label{elmergreen} B.G. Elmergreen and Y.N. Efremov, 
\apj {\bf 480}, 235 (1997).

\item \label{andromeda}  E. Brinks and E. Bajaja, 
\aa {\bf 169}, 14 (1986).

\item \label{orientation} It is important to stress the process described 
in this report can be very effective at modifying galactic morphologies
throughout clusters.
Following the results in (\ref{ghigna}) who determined the average galaxy orbit 
in clusters, we have estimated that more than 90\% 
of galaxies within a rich virialised 
cluster can be completely stripped of their diffuse HI. 
This calculation relied on the 
facts that: i) the typical time scale of the stripping is very short compare
with the orbital characteristic time of a galaxy in cluster 
$\sim10^8$ years, ii) even for galaxies moving with a relative angle of 
20 degrees to the ICM, all the gas is stripped (see Figure 1), and 
iii) the form of the galactic orbits 
-- typically with pericenters less than 500 kpc and an average relation
apocenter:pericenter approximately 6:1 -- 
in clusters and all possible orientations respect the ICM. 

\item \label{vollmer} B. Vollmer, V. Cayatte, A. 
Boseli, C. Balkowski, W.J. Duschl, 
\aa {\bf 349}, 411 (1999).

\item \label{ryder} S.D. Ryder, G. Purcell, D. Davis, V. Anderson, 
\pasp {\bf 14}, 1 (1997).

\item \label{radiojet}
Indirect evidence for cluster-galaxy hydro-dynamic interactions is frequently
observed in head-tail radio galaxies with escaping radio-jets that are 
stripped
backwards by the ICM. This long lived synchrotron emission may arise within the
stripped and subsequently ionised plasma that is in pressure equilibrium with
the ICM, supported by the original 
disk magnetic field that is entangled with the
stripped disk material (see Figure 4 in http://www.scienceonline.org). 

\item \label{warmels} R.H. Warmels, 
 \aass {\bf 72}, 57 (1988).

\item \label{cayatte}  V. Cayatte, C. Balkowski, J.H.  
Van Gorkom, C. Kotanyi, 
\aj {\bf 100}, 604 (1990).

\item \label{bravo} H. Bravo-Alfaro, V. Cayatte, 
J.H. van Gorkom, C. Balkowski,  \aj {\bf 119}, 580 (2000). 

\item \label{blitz} L. Blitz and F.H. Shu , \apj, {\bf 238}, 148 (1980).

\item \label{rgbnote}
If the lifetimes of dense molecular clouds are as short as $10^7$ years, 
then the decline in the molecular gas content of infalling galaxies is
as rapid as the rate of HI removal by the stripping process. Initially,
this seems at odds with the observations of J.D.P. Kenney and  J.S. Young 
, \apj, {\bf 344}, 171 (1989), who found that bright HI deficient spirals
in the Virgo cluster contained similar masses of molecular hydrogen to
counter-parts of the same morphological type in the field. This
apparently suggests that molecular clouds must have a lifetime that
is considerably longer than the stripping timescale. However, we note
that this comparison is made at a fixed morphology. Morphology
is strongly dependent on the star formation rate, in the sense that 
galaxies with low star formation rate will be classified as earlier type. 
Thus it is unlikely that galaxies with similar morphology will exhibit
large differences in CO content. Rather a large deficiency in CO will
result in a galaxy with low star formation rate, and earlier
morphological type. It is then hard to disentangle any deficiency in 
CO content due to stripping from the reduction in CO content expected 
for the change in morphological type. It is encouraging, 
nevertheless, that galaxies
of earlier type match more closely the curve  in Kenney and  Young's data
expected 
if the molecular and atomic gas contents decline at similar rates.

\item \label{fujita} Y. Fujita and M. Nagashima, 
\apj {\bf 516},  619 (1999).

\item \label{gavazzi} G. Gavazzi et al.,
\aa {\bf 304}, 325 (1995).

\item \label{kenney}J.D.P. Kenney and R.A. Koopmann, 
\aj {\bf 117}, 181 (1999).

\item VQ is a Marie Curie research fellow of the European Union
 (grant HPMF-CT-1999-00052).
During the first part of this work, VQ was supported by 
a fellowship of the
Spanish SEUID (Ministerio de Educaci\'on y Cultura) 
and partially by Spanish DGES (grant PB96-0797).
BM is a Royal Society research fellow.
Numerical simulations
were carried out 
as part of the Virgo Consortium and 
the UK Computational Cosmology Consortium.

\end{enumerate}

\newpage

\bigskip

\noindent{\bf FIGURES CAPTIONS}

\bigskip

\noindent{\bf Figure 1} { \ \ The evolution of the gaseous disk of a spiral
galaxy moving face on (left column) and inclined 20 degrees 
to the direction of motion (right
column)
through a diffuse hot intra-cluster medium.  Each snapshot
shows the density of gas ($\delta=\rho/\rho_{_{ICM}}$) within a 0.2 
kpc slice through the center of the galaxy
and each frame is 64 kpc on a side.  Note how rapidly the disk material is
removed - within 100 million years 100\% of the HI is lost.  We do not 
show the
stellar disk, bulge or dark matter halo which remain unaffected by the loss of
the gaseous component. The box size is 64 kpc and the hydro grid has 
$256^3$ cells. 

\bigskip

\noindent{\bf Figure 2} { \ \ Mass loss as a function of time for 
the model including ten little holes and inhomogeneous density.
We plot
the evolution of the gaseous mass within a cylindrical slice of 25 kpc 
radius and 
thickness 2 kpc  centred on the center of mass of the stellar disk. 
Initially, ram-pressure stripping dominates the gas loss process and the
entire outer disk is removed in a very rapid timescale. 
Viscous and turbulent stripping operates continuously, but over a longer
timescale, resulting in a roughly linear rate of mass loss. 
We stop this simulation after 130 million years by which time 97\% of
the gas disk has been removed.} 

\bigskip

\noindent{\bf Figure 3} { \ \ Observational evidence (left panels) for
ram-pressure processes compared with our hydro-dynamical simulations at 
different
epochs (right panels).  The first panel shows a HI map of NGC 7421 which shows
wings of gas being pushed back by its motion through a diffuse ionised medium
(\ref{ryder}).  The second panel shows the HI deficient galaxy NGC 4548
orbiting in the Virgo cluster (\ref{vollmer}).  The remaining gas has a
ring-like morphology very similar to our simulation after 50 million years. 

\bigskip

\begin{center}
{\bf Next Figures would be shown in http://www.scienceonline.org as 
an extra material}
\end{center}

\bigskip

\noindent{\bf Figure 4} { \ \ Observational evidence (left panels) of
trails of gas compared with our simulations at different epochs (right 
panels). The first panel shows the radio continuum (contours) brightness 
distribution superposed over a gray scale representation of H intensity 
in galaxy 97073 in cluster A 1367 (see Figure 1a in \ref{gavazzi}). 
The second panel shows spectacular radio jets that have been swept 
backwards for tens of kpc by motion of the galaxy through the ICM
of a rich cluster at z=0.3 (radio data courtesy of 
R. Ivison, A.W. Blain and I. Smail). 
Note that the radio emission may be unrelated to the stripping
process, but the morphological appearance is very similar to the
trails of stripped gas in our simulations.}

\end{document}